\documentclass[a4paper]{jpconf}
\usepackage{dcolumn}% Align table columns on decimal point
\usepackage{bm}% bold math
\usepackage{verbatim}
\usepackage[english]{babel}
\usepackage{amsmath}
\usepackage{amsfonts}
\usepackage{pifont}
\usepackage[T1]{fontenc}

\usepackage{amssymb}
\usepackage{wasysym}
\usepackage{hyperref}
\usepackage{enumerate}
\usepackage{amsmath}
\usepackage{amsfonts}
\usepackage{amssymb}
\usepackage{amssymb}
\usepackage{wasysym}
\usepackage{hyperref}
\usepackage{enumerate}
\usepackage{wrapfig}%per fare le figure con testo accanto si usa con ambiente 
\usepackage{graphicx}
\usepackage{color}
\usepackage{rotating}
\usepackage{eso-pic}
\usepackage[footnotesize,bf]{caption}

\usepackage{subfigure}
\usepackage{eso-pic}
\usepackage{bm}% bold math

\newcommand{\bq}    {\begin{equation}}
\newcommand{\eq}    {\end{equation}}
\newcommand{\bqr} {\begin{eqnarray}}
\newcommand{\eqr} {\end{eqnarray}}

\begin{document}

\title{A hybrid-chiral soliton model with broken scale invariance for nuclear matter}

\author{A. Drago$^1$, V. Mantovani Sarti$^1$}

\address{$^1$Department of Physics, University of Ferrara, 
ITALY}

\ead{drago@fe.infn.it, smantovani@fe.infn.it}

\begin{abstract}

We present a model for describing nuclear matter at finite density based on quarks interacting with chiral fields, $\sigma$ and $\pi$.
 The chiral Lagrangian also includes a logarithmic potential, associated with the breaking of scale invariance. We provide results for the soliton in vacuum and at finite density, using the Wigner-Seitz approximation. We show that the model can reach higher densities respect to the Linear-$\sigma$ model, up to $\rho \approx 3 \rho_0$ for $m_{\sigma}=1200$ MeV.

\end{abstract}

\section{Introduction}
%approccio Carter VS approccio Serot (diversa introduzione campo scalare)\\
%Quark model in nuclear physics\\
%Accenno a Pirner (Non local NJL) e Thomas (Local NJL+confinamento) \\
%Wigner-Seitz approx, citare lavori precedenti (Arturino,Giuditta, spagnoli)\\

The phase diagram of nuclear matter can be in principle more complicated than the one based on MIT-bag-like models which
predict a direct transition from hadronic matter to quark-gluon plasma (QGP). 
In the region of high densities and relatively low temperatures a variety of phases can exist in which
chiral-symmetry breaking is realized in different ways. 
On the other hand, to study systems at finite density by using chiral lagrangians is not a trivial task: for instance models based on the Linear $\sigma$-model fail to describe nuclear matter already at $\rho \sim \rho_0$. In Ref.~\cite{R.J.Furnstahl1993} the authors
conclude that the failure of the $\sigma$ model
is due to the restrictions on the scalar field dynamics imposed by the Mexican
hat potential.
In Ref.~\cite{R.J.Furnstahl1996} they use a non-linear realization of chiral symmetry in which a scalar-isoscalar effective field is introduced, as a chiral singlet, to simulate
intermediate range attraction.
In this way the dynamics of the chiral singlet field is no more regulated by the
Mexican hat potential and accurate results for finite nuclei can be found.\\

Another possible solution to the problem of studying systems at finite density in chiral models is to use a linear realization but with a new potential, which includes terms not present in the Mexican hat potential.
A possible guideline in building such a potential is scale invariance.

In QCD scale invariance is spontaneously broken due to the presence of the parameter $\Lambda _{QCD}$ coming from the renormalization process.
Formally, the non conservation of the dilatation current is strictly connected to a not vanishing gluon condensate \cite{E.K.Heide1994}, \cite{G.W.Carter1998}, \cite{G.W.Carter1997}, \cite{G.W.Carter1996}:
\begin{equation}\label{current}
\langle \partial_{\mu}j^{\mu}_{QCD} \rangle =\dfrac{\beta (g)}{2g}\langle F^{a}_{\mu \nu}(x)F^{a\mu \nu}(x) \rangle .
\end{equation}

In the approach of Schechter, Migdal, and Shifman \cite{J.Schechter1980}
a scalar field representing the gluon condensate is introduced and its dynamics is regulated by a potential chosen so that it reproduces (at mean-field level) the divergence of the scale current that in QCD is given by Eq.~(\ref{current}). The potential of the dilaton field is therefore determined by the equation:
\begin{equation}\label{dil}
\theta_\mu ^\mu = 4 V(\phi)-\phi \dfrac{\partial V}{\partial \phi}=4 \epsilon_{vac} \left (\dfrac{\phi}{\phi_0}\right )^4
\end{equation}
 where the parameter $ \epsilon_{vac}$ represents the vacuum energy.
To take into account massless quarks a generalization was proposed in Ref.~\cite{E.K.Heide1992}, so that also
chiral fields contribute to the trace anomaly. In this way the single scalar field of Eq.~(\ref{dil}) is replaced by a set of scalar fields
$\lbrace \sigma, \boldsymbol{\pi}, \phi \rbrace$.

It
has already been shown that an hadronic model based on this dynamics provides a good
description of nuclear physics at densities about $\rho_0$ and descibes the gradual restoration of chiral symmetry at higher densities~\cite{Bonanno:2008tt}.

The new idea we develop in this work is to interpret the fermions as quarks, to build the hadrons as solitonic solutions of the fields equations and finally to explore the properties of the soliton at finite density using the Wigner-Seitz approximation.\\

The structure of the paper is the following. In Sec.~\ref{model} we describe the model we are using, in Sec.~\ref{results} we discuss the results for the single soliton, first in vacuum and then at finite density. Finally, in Sec.~\ref{conclusions} we present our conclusions and future outlooks.

\section{The model}\label{model}
The basic approach of a hybrid chiral model is to couple quarks to mesons in a chirally invariant way. We consider such a Lagrangian:
\begin{equation}\label{lagr}
\mathcal{L}  =\bar{\psi}  \big(i \gamma ^\mu\partial _{\mu}-g_{\pi}(\sigma + i \boldsymbol{\pi}\cdot \boldsymbol{\tau}\gamma _{5}))\psi 
  +\dfrac{1}{2}(\partial_{\mu}\sigma \partial^{\mu}\sigma + \partial_{\mu}\boldsymbol{\pi}\cdot \partial^{\mu}\boldsymbol{\pi})-V(\phi_0,\sigma,\mathbf{\pi}).
\end{equation}

Here $\psi$ is the quark field, $\sigma$ and $\pi$ are the chiral fields and $\phi$ is
the dilaton field which, in the present calculation, is kept frozen at its vacuum value $\phi _0$. 
The potential is given by:
\begin{eqnarray} \label{pot}
V(\phi_0,\sigma,\mathbf{\pi})& = & -\dfrac{1}{4}B\phi_0^4 -\dfrac{1}{2}B \delta \phi_0^4 Log \dfrac{\sigma^2 + 
\boldsymbol{\pi}^2}{\sigma_{0} ^2}\nonumber\\
& & +\dfrac{1}{2}B\delta \dfrac{\phi^4_0}{\sigma^2_0} 
\left (\sigma^2 + \boldsymbol{\pi}^2-\dfrac{\sigma^2_0}{2}\right )\nonumber \\
& & - \dfrac{1}{4} \epsilon_{1} \left[\dfrac{4\sigma}{\sigma _{0}}
 -2 \left (\dfrac{\sigma^2 + \boldsymbol{\pi}^2}{\sigma_0^2}\right)+2\right]\nonumber \\
 & & -\dfrac{1}{4}B\phi^4_0 (\delta -1)+\epsilon_{1}
\end{eqnarray}
where the logarithmic term generates from~(\ref{dil}).
The constants in the last line of Eq.~(\ref{pot}) ensure that the vacuum energy is zero.\\
The constants $B$ and $\phi_0$ can be fixed by choosing value for the mass of the glueball and the vacuum energy $\epsilon_{vac}$, while $\delta=4/33$ is provided by the QCD beta function and it corresponds to the relative weight of the fermionic and of the gluonic degrees of freedom.

The first two terms of the potential are responsible for the breaking of scale invariance, while the term in the third line explicitly breaks the chiral invariance of the lagrangian.
The Euler-Lagrangian equations that follow from Lagrangian~(\ref{lagr}) are:

\begin{align}
&[i\displaystyle{\not} \partial -g_{\pi}(\sigma + i \boldsymbol{\pi}\cdot \boldsymbol{\tau}\gamma _{5})]\psi = 0 \nonumber,\\
& \partial _\mu \partial ^\mu \sigma = -g\bar{\psi}\psi -\dfrac{\partial V}{\partial \sigma}\nonumber,\\
& \partial _\mu \partial ^\mu \boldsymbol{\pi} = -i g \bar{\psi}\boldsymbol{\tau}\gamma_5 \psi-\dfrac{\partial V}{\partial \boldsymbol{\pi}}.
\end{align}
\normalsize

The values of the parameters used for calculations in vacuum are listed in Table~\ref{tab}.\\
\newline
\begingroup
%\squeezetable
\begin{table}[h]
\caption{ Values of the parameters.}\label{tab}
\centering
\begin{tabular}{|c|c|}
\hline  Quantity&Value  \\ 
\hline  $\vert\epsilon_{vac}\vert ^{1/4}$ (MeV)& 236 \\ 
\hline  g & 5  \\ 
\hline  $m_\sigma$ (MeV)& 550  \\ 
\hline  $B$ & 24.37\\
\hline  $\epsilon_1^{1/4}$ (MeV)&  114\\ 
\hline  $\delta$ & $4/33$  \\ 
\hline  $\sigma_0$ (MeV)& 93  \\ 
\hline  $\phi_0$ (MeV)& 175.23\\
\hline 
\end{tabular} 
\end{table}
\endgroup
\subsection{The hedgehog ansatz}
We are working in the mean-field approximation, where mesons are described by time-independent, classical fields and where powers and products of these fields are replaced by powers and products of their expectation values.
The quark spinor in the spin-isospin space is:
\begin{equation}
\psi=\dfrac{1}{\sqrt{4 \pi}}\left (\begin{array}{c}
u(r)\\
iv(r)\boldsymbol{\sigma} \cdot \boldsymbol{\hat r}\end{array}\right) \dfrac{1}{\sqrt 2}(\vert u\downarrow\rangle-\vert d\uparrow\rangle)
\end{equation}
Once fixed the form of the quark wave function the self-consistents ansatze for the meson fields are:
\small
\begin{align}
& \langle \hat{\sigma} \rangle =\sigma (r)\nonumber ,\quad \langle \boldsymbol{\hat{\pi}}_a\rangle =r_a \pi (r)
\end{align}
\normalsize
 where $\sigma (r)$, $\pi (r)$ are radial functions of $r$. \\

\section{Results}\label{results}
\subsection{Single soliton in vacuum}

Solving the vacuum case requires the following boundary conditions for the fields:

\begin{eqnarray}\label{bcvac}
& u'(0)=v(0)=0\nonumber ,\\
&\sigma '(0)=\pi (0)=0
\end{eqnarray}
while at infinity (in practice at a value $r=R=4$ fm), the boundary conditions read:
\small
\begin{align}
& \sigma (R)=\sigma_0 ,\,  \pi (R)=0,\nonumber\\
& \dfrac{v(R)}{u(R)}=\sqrt{\dfrac{-g \sigma (R)+\varepsilon}{-g \sigma (R)-\varepsilon}}.
\end{align}
\normalsize
A test for convergence of the solution comes from another way of expressing the energy obtained by Rafelski~\cite{Rafelski1977} by integrating out the fermionic fields:
 \begin{align}
E_{Raf.}&=\int _a ^R d^3r\bigg[4\left(V-\sigma\dfrac{\partial V}{\partial \sigma}-\pi \dfrac{\partial V}{\partial \pi}\right)\bigg].
\end{align}
Our solutions satisfy this consistency test up to a precision of the order of $10^{-3}$.\\

 In Table~\ref{res} we present the static properties of the nucleon at Mean Field level and we compare them with experimental values and 
in Table~\ref{contrib} we show the decomposition of the soliton total energy in various contributions and the comparison with the Linear-$\sigma$ model~\cite{Birse1985}.\\
\normalsize
 \begin{figure}[]
\centering
\includegraphics[width=\textwidth]{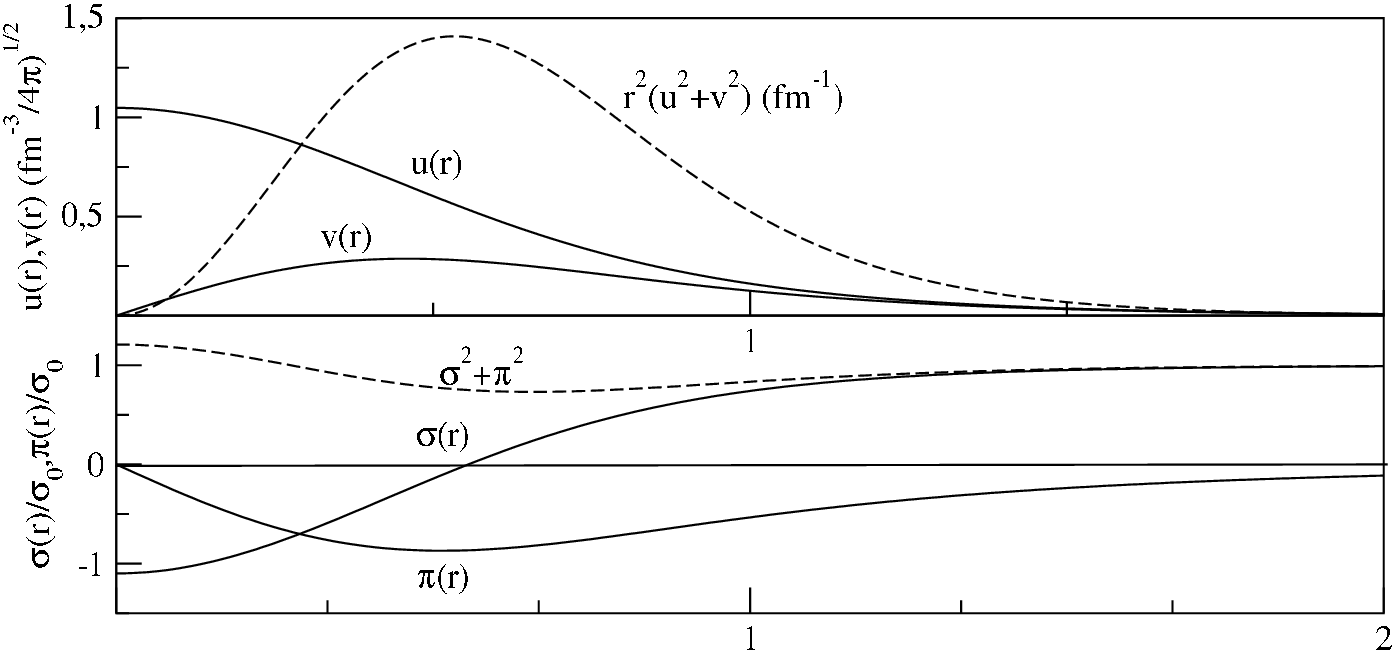} 
\caption{Fields in vacuum for the solution $g=5, m_\sigma =550$ MeV.}\label{vacuum}
\end{figure}

\begingroup
%\squeezetable
\begin{table}[h!]
\caption{Various nucleon properties at mean field level in the present work and comparison with experimental values.}\label{res}
\centering
\begin{tabular}{|c | c  c|}
\hline
 Quantity & MFA & Exp.  \\
 \hline
 $M\,(MeV)$ & $1175.6$  & $1085$\\[5pt]
 $\langle r_e ^2\rangle_{I=0}$& $(0.73\, fm)^2$  & $(0.72\, fm)^2$ \\ [5pt]
 $\langle r_m ^2\rangle_{I=1}$& $(0.91\, fm)^2$ & $(0.80\, fm)^2$ \\[5pt] 
% $\mu_{I=1}$ $(\mu_N)$& $3.83$ & $4.70$ \\[5pt] 
 $g_A$ & $1.27$   & $1.26$ \\ [10pt]
 \hline
\end{tabular} 
\end{table} 
\endgroup

\begingroup
%\squeezetable
\begin{table}[h!]
\caption{Contributions to the soliton total energy in the Logarithmic model and comparison with Linear-$\sigma$ model. All quantities in MeV.}\label{contrib}
\centering
\begin{tabular}{|c c c|c|c|}
\hline
Quantity &  &  & Log. Model & Linear-$\sigma$ Model  \\
\hline
Quark eigenvalue &  &  & $83.1$ & $107.4$\\[5pt]
Quark kinetic energy &  &  & $1138$& $1056.9$ \\ [5pt] 
$E_\sigma$ (mass+kin.)&  &  & $334.5$& $320.3$ \\[5pt]
$E_\pi$ (mass+kin.)&   &  & $486$ & $373.1$\\ [5pt]
Potential energy $\sigma -\pi$& & & $105.7$& $120.7$ \\ [5pt]
$E_{q\sigma}$&   &   & $-101.4$ & $-62.3$\\ [5pt]
$E_{q\pi}$ &   &  & $-787$ & $-673.2$\\ [5pt]
Total energy  &   &  & $1175.6$ & $1136.2$\\ [10pt]  
 \hline
\end{tabular} 
\end{table}
\endgroup

%\begingroup
%%\squeezetable
%\begin{table}[h!]
%\caption{Various nucleon properties at MF level in the present work and comparison with experimental values.}\label{res}
%\centering
%\begin{tabular}{|c | c  c|}
%\hline
% Quantity & MFA & Exp.  \\
% \hline
% $M\,(MeV)$ & $1175.6$  & $1085$\\[5pt]
% $\langle r_e ^2\rangle_{I=0}$& $(0.73\, fm)^2$  & $(0.72\, fm)^2$ \\ [5pt]
% $\langle r_m ^2\rangle_{I=1}$& $(0.91\, fm)^2$ & $(0.80\, fm)^2$ \\[5pt] 
%% $\mu_{I=1}$ $(\mu_N)$& $3.83$ & $4.70$ \\[5pt] 
% $g_A$ & $1.27$   & $1.26$ \\ [10pt]
% \hline
%\end{tabular} 
%\end{table} 
%\endgroup

\newpage
\subsection{Single soliton at finite density}

 In order to describe a soliton system at finite density we use the Wigner-Seitz approximation. This approach is very common for this kind of calculations and it has already been widely applied to both non-linear~\cite{D.Hahn1987,N.K.Glendenning1986,P.Amore2000} and Linear-$\sigma$ models~\cite{U.Weber1998}.\\
Specifically the Wigner-Seitz approximation consists of replacing the cubic lattice by a spherical
symmetric one where each soliton sits on a spherical cell of radius R with
specific boundary conditions imposed on fields at the surface of the sphere.
The specific configuration of the meson fields, each one centered at a lattice point, generates a periodic potential in which the quarks move.\\
In particular now the spinor of quark fields must satisfy the Bloch theorem:

\begin{equation}
\psi _{\boldsymbol{k}} (r)=e^{i \boldsymbol{k}\cdot \boldsymbol{r}}\Phi_{ \boldsymbol{k}}(r),
\end{equation}
where $\boldsymbol{k}$ is the crystal momentum (for the ground state is equal to zero) and $\Phi_{ \boldsymbol{k}}(r)$ is a spinor that has the same periodicity of the lattice.\\
We should explain in detail our choice of boundary conditions, because there are various sets of possible boundary conditions~\cite{U.Weber1998,P.Amore2000}.
In particular we relate the choice of our boundary conditions to the parity $\boldsymbol{r}\rightarrow -\boldsymbol{r}$ operation; respect to this symmetry the lower component $v(r)$ of quark spinor, the pion $\pi (r)$ and the $\rho (r)$ fields are odd, so all of these fields should vanish at R:
\begin{equation}
v(R)=\pi (R)=\rho (R)=0.
\end{equation}
For the remaining fields and the upper Dirac component we apply the usual "flatness" condition on the boundary:
\begin{equation}
u'(R)=\sigma '(R)=\omega '(R)=A_S'(R)=A_T'(R)=0.
\end{equation}

Basically the calculation is based on solving the set of coupled fields equations in a self-consistent way for each step in R; we start from the vacuum value, $R=4$ fm, and we slowly decrease the cell radius down to the smallest radius for which self-consistent solutions can be obtained.
\newpage

\begin{figure}[!h]
\centering
 \includegraphics[scale=0.5]{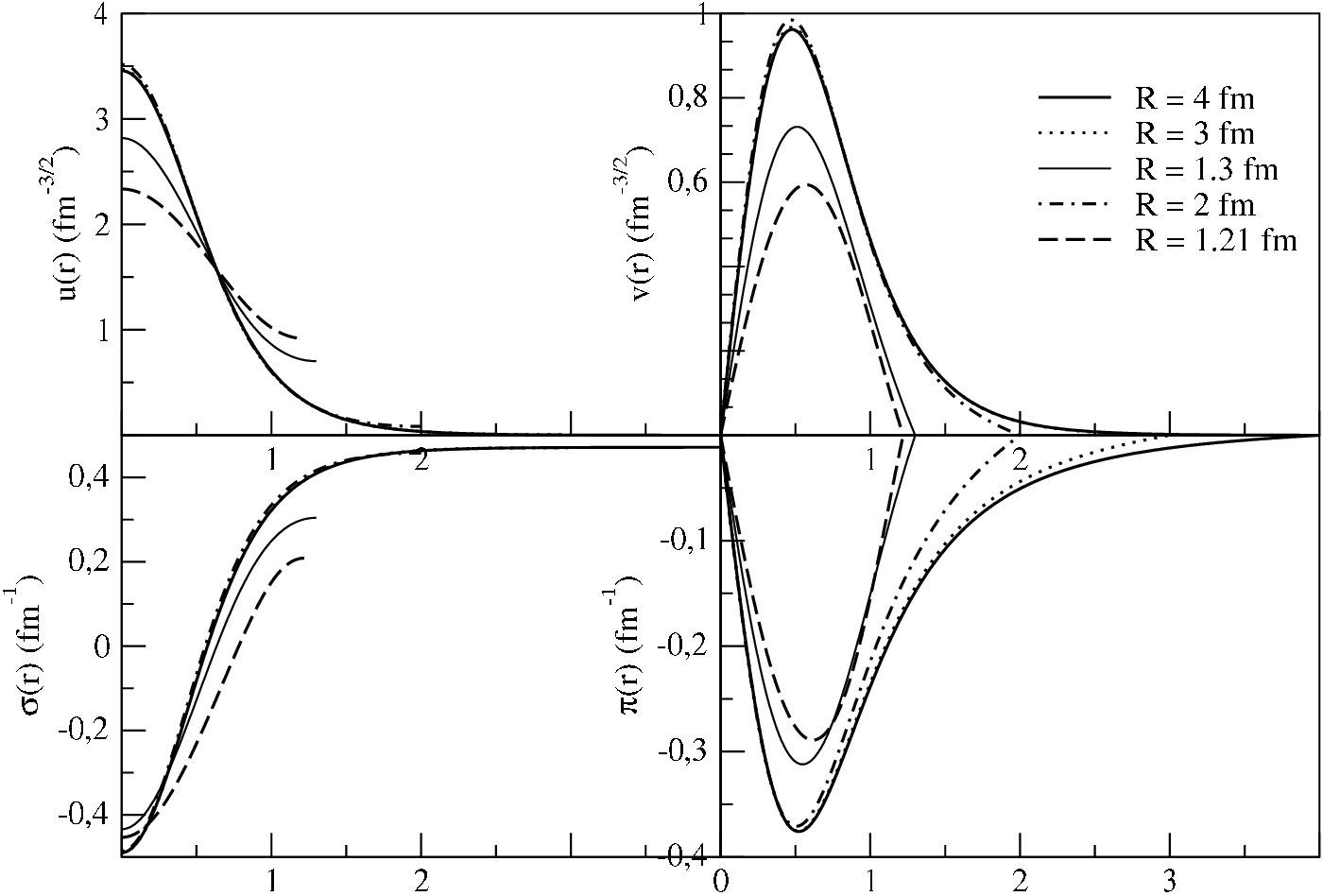}
\caption{Upper and lower components of the Dirac spinor, $\sigma$ and pion fields for different values of the cell radius $R$.}\label{campi}
\end{figure}

\begin{figure}
\centering
\includegraphics[width=\textwidth]{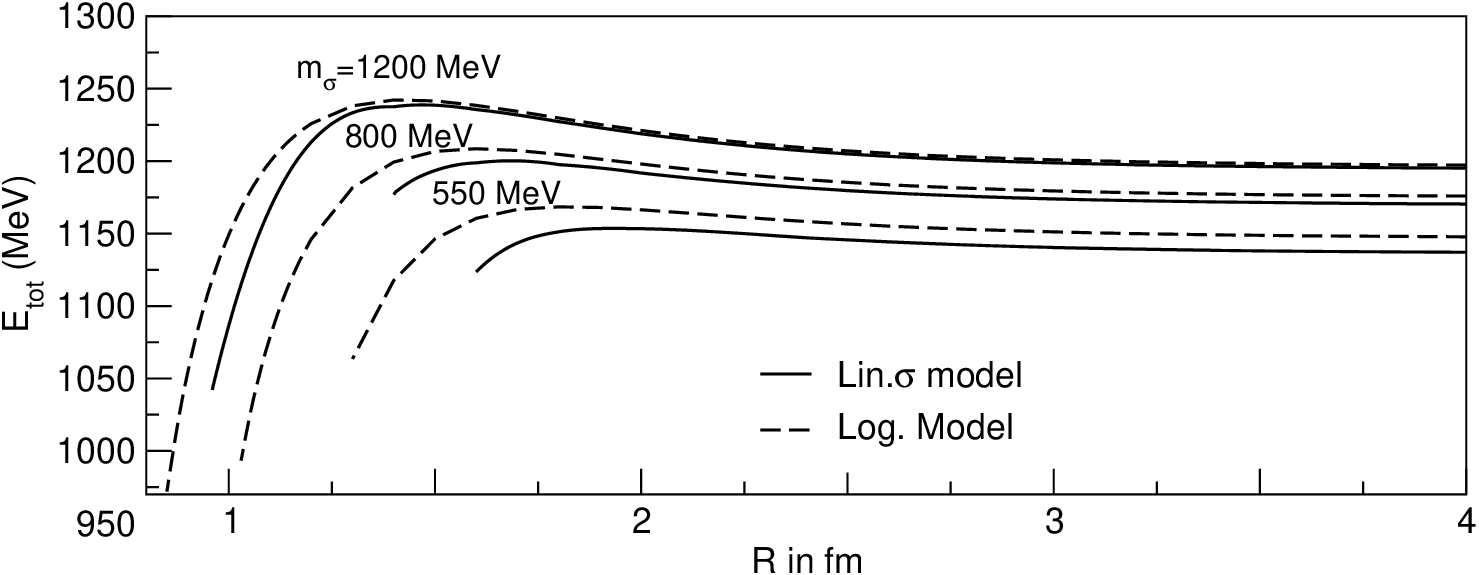} 
\caption{Total energy of the soliton as a function of cell radius $R$ for the Linear-$\sigma$ model \cite{U.Weber1998} for different values of $m_\sigma$ and present model.}\label{Etot}
\end{figure}

%\begin{figure}[!t]
%\centering
%\includegraphics[clip,width=0.6\textwidth]{campi_densfinita.eps} 
%\caption{Upper and lower components of the Dirac spinor, $\sigma$ and pion fields for different values of the cell radius $R$.}\label{campi}
%\end{figure}
In Fig.~\ref{campi} we plot the Dirac and the chiral fields for different values of $R$; up to $R=2$ fm, the solutions do not change significantly, but as the cell radius shrinks to lower values, we see that all the fields are deeply modified by the finite density.
In Fig.~\ref{Etot} we show the results, for the total energy of the soliton, in the present model and in the Linear-$\sigma$ model.
For a fixed value of the $\sigma$ mass, the logarithmic model do reach higher densities; as the $\sigma$ mass raises, the system can get to lower $R$ because chiral fields are stuck on the chiral circle.

\newpage

\section{Conclusions}\label{conclusions}
We used a Lagrangian with quarks degrees of freedom based on chiral and scale invariance to study how the soliton behaves in vacuum and at finite density.
At zero density the interplay between quarks and chiral fields lead to a soliton of mass $M=1175.6 MeV$, not too far from the experimental value given by the medium value $1085$ MeV of $M_n+M_\Delta$. We employed the Wigner-Seitz approximation to describe the dense system and we compared our results to the Linear-$\sigma$ model~\cite{U.Weber1998}.
In this approach we showed that the new potential, which includes the scale invariance, allows the system to reach higher densities at fixed $m_{\sigma}$.
This improvement is totally due to the different dynamics between chiral fields, given by the logarithmic potential.\\

The present work will be extended in several directions. First of all the finite density approach will be applied to a model including vector mesons, in order provide the necessary repulsion at large density. Next we will provide a precise and accurate calculation of the eigenvalues band in the soliton crystal in order to see how the system is affected by these effects~\cite{U.Weber1998}.
 Finally, the model can also be studied also at finite temperature, including the dynamics of the dilaton field. 
It is interesting to note that when this model has been investigated at finite temperature assuming the fermions to be hadrons~\cite{Bonanno:2008tt}, a phase diagram similar to the one proposed by McLerran and Pisarski~\cite{L.McLerran2007} was obtained. 
In principle the approach based on the Wigner-Seitz scheme should be able to recover a scenario similar to the one discussed in~\cite{Bonanno:2008tt}, but starting from more fundamental ingredients.

\ack

It is a pleasure to thank B.Y.Park and V. Vento for many stimulating discussions, M.Birse and J.McGovern for useful comments and tips on calculations.

\section*{References}

\bibliographystyle{iopart-num}
\bibliography{biblio}

\providecommand{\newblock}{}
\begin{thebibliography}{10}
\expandafter\ifx\csname url\endcsname\relax
  \def\url#1{{\tt #1}}\fi
\expandafter\ifx\csname urlprefix\endcsname\relax\def\urlprefix{URL }\fi
\providecommand{\eprint}[2][]{\url{#2}}

\bibitem{R.J.Furnstahl1993}
Furnstahl R~J and Serot B~D 1993 {\em Phys. Rev.\/} {\bf C47} 2338--2343

\bibitem{R.J.Furnstahl1996}
Furnstahl R~J, Serot B~D and Tang H~B 1996 {\em Nucl. Phys.\/} {\bf A598}
  539--582 (\textit{Preprint} \eprint{nucl-th/9511028})

\bibitem{E.K.Heide1994}
Heide E~K, Rudaz S and Ellis P~J 1994 {\em Nucl. Phys.\/} {\bf A571} 713--732
  (\textit{Preprint} \eprint{nucl-th/9308002})

\bibitem{G.W.Carter1998}
Carter G~W and Ellis P~J 1998 {\em Nucl. Phys.\/} {\bf A628} 325--344
  (\textit{Preprint} \eprint{nucl-th/9707051})

\bibitem{G.W.Carter1997}
Carter G~W, Ellis P~J and Rudaz S 1997 {\em Nucl. Phys.\/} {\bf A618} 317--329
  (\textit{Preprint} \eprint{nucl-th/9612043})

\bibitem{G.W.Carter1996}
Carter G~W, Ellis P~J and Rudaz S 1996 {\em Nucl. Phys.\/} {\bf A603} 367--386
  (\textit{Preprint} \eprint{nucl-th/9512033})

\bibitem{J.Schechter1980}
Schechter J 1980 {\em Phys. Rev.\/} {\bf D21} 3393--3400

\bibitem{E.K.Heide1992}
Heide E~K, Rudaz S and Ellis P~J 1992 {\em Phys. Lett.\/} {\bf B293} 259--264

\bibitem{Bonanno:2008tt}
Bonanno L and Drago A 2009 {\em Phys.Rev.\/} {\bf C79} 045801
  (\textit{Preprint} \eprint{0805.4188})

\bibitem{Rafelski1977}
Rafelski J 1977 {\em Phys. Rev.\/} {\bf D16} 1890

\bibitem{Birse1985}
Birse M~C and Banerjee M~K 1985 {\em Phys.Rev.\/} {\bf D31} 118

\bibitem{D.Hahn1987}
Hahn D and Glendenning N~K 1987 {\em Phys. Rev.\/} {\bf C36} 1181

\bibitem{N.K.Glendenning1986}
Glendenning N~K 1986 {\em Phys. Rev.\/} {\bf C34} 1072--1080

\bibitem{P.Amore2000}
Amore P and De~Pace A 2000 {\em Phys. Rev.\/} {\bf C61} 055201
  (\textit{Preprint} \eprint{nucl-th/9910074})

\bibitem{U.Weber1998}
Weber U and McGovern J~A 1998 {\em Phys. Rev.\/} {\bf C57} 3376--3383
  (\textit{Preprint} \eprint{nucl-th/9710021})

\bibitem{L.McLerran2007}
McLerran L and Pisarski R~D 2007 {\em Nucl. Phys.\/} {\bf A796} 83--100
  (\textit{Preprint} \eprint{0706.2191})

\end{thebibliography}

\end{document}